\documentclass[prd,aps,nofootinbib,11pt]{revtex4}
\usepackage{graphicx,epsfig,color}

\begin{document}
\title{Test Flavor SU(3) symmetry in Exclusive $\Lambda_c$ decays }
\author{Cai-Dian L\"u$^{1}$~\footnote{Email:lucd@ihep.ac.cn}, Wei Wang $^{2,3}$~\footnote{Email:wei.wang@sjtu.edu.cn} and Fu-Sheng Yu$^4$~\footnote{Email:yufsh@lzu.edu.cn} }
\affiliation{
$^1$ Institute of High Energy Physics, P.O. Box 918(4), Chinese Academy of Sciences, Beijing 100049, People's Republic of China \\
$^2$ INPAC, Shanghai Key Laboratory for Particle Physics and Cosmology, Department of Physics and Astronomy, Shanghai Jiao-Tong University, Shanghai, 200240,   China\\
$^3$ State Key Laboratory of Theoretical Physics, Institute of Theoretical Physics, Chinese Academy of Sciences, Beijing 100190, China\\
$^4$ School of Nuclear Science and Technology, Lanzhou University, Lanzhou 730000, People's Republic of China
}

\begin{abstract}
Flavor SU(3) symmetry is a powerful  tool to analyze   charmed baryon decays, however its  applicability  remains  to be experimentally validated.  Since there is not much  data on $\Xi_c$ decays, various exclusive $\Lambda_c$ decays especially  the ones into a neutron state are  essential  for the test of flavor symmetry. These decay modes are also helpful to investigate  final state interactions in charmed baryon decays. In this work, we discuss  the  explicit roles of $\Lambda_c$ decays into a neutron in testing the flavor symmetry  and exploring  final state interactions. The involved  decay modes  include   semileptonic decays,  two-body and three-body non-leptonic decays, but  all of them  have not been experimentally observed to date. 
\end{abstract}
\maketitle

\section{Introduction}

Charmed baryon decays, in particular   $\Lambda_c$ and $\Xi_c$ decays,  are of great interest as they serve as a platform for the study of strong and weak interactions in  heavy-to-light baryonic transitions. They  can also provide  the essential  inputs for the  $\Lambda_b$ decay modes into  a charmed baryon like $\Lambda_c$.  On the experimental side, most  available results on $\Lambda_c$ decays are obtained using the old data until   recently.  In 2014, Belle collaboration  provided an measurement  of the branching fraction with a very small uncertainty~\cite{Zupanc:2013iki}, 
\begin{equation}
{\cal B}(\Lambda_c^+ \to p K^- \pi^+)_{\rm Belle}=(6.84\pm0.24^{+0.21}_{-0.27})\%, \label{eq:belle_pKpi}
\end{equation}
but  the central value is much larger than the previous measurement by the CLEO-c collaboration~\cite{Jaffe:2000nw}:
\begin{eqnarray}
{\cal B}(\Lambda_c^+ \to p K^- \pi^+)_{\rm CLEO}=(5.0\pm0.5\pm1.2)\%. \label{eq:cleo_pKpi}
\end{eqnarray} 
Based on the large amount of data, Belle collaboration   also started to study  the doubly Cabibbo-suppressed processes~\cite{Yang:2015ytm}.   Making use of the data collected  in the $e^+e^-$ collision at the center-of-mass energy of $\sqrt s=4.599$ GeV and adopting  the double-tag technique, BES-III  collaboration has reported  first measurements of absolute hadronic branching fractions of  Cabibbo-favored decay modes~\cite{Ablikim:2015flg}. In total, twelve $\Lambda_c$ decay modes were observed  with the significant improvement on the branching fraction in particular for the $\Lambda_c\to pK^-\pi^+$:
\begin{eqnarray} 
{\cal B}(\Lambda_c^+\to pK^-\pi^+)_{\rm BESIII} = (5.84\pm 0.27\pm 0.23)\%. \label{eq:bes_pKpi}
\end{eqnarray}
While the uncertainties are comparable with the Belle results in Eq.~(\ref{eq:belle_pKpi}), its  central value is much smaller, which is closer to the central value of the CLEO results in Eq.~(\ref{eq:cleo_pKpi}).   We believe this difference will be clarified in future since  the experimental prospect on charmed baryon decays will be very promising~\cite{Asner:2008nq,Aushev:2010bq}.

Theoretical description of charmed baryon decays is mostly  based on the factorization assumption together with the analysis of some  non-factorizable contributions in nonperturbative  explicit modes~\cite{Cheng:1991sn,Zenczykowski:1993hw,Uppal:1994pt,Fayyazuddin:1996iy}. However  the factorization scheme does not seem to be supported by experiments, for instance the observed large branching fraction for decays like $\Lambda_c\to \Sigma^+\pi^0/\Xi^0K^+$, which are forbidden in the factorization scheme~\cite{Agashe:2014kda}. An alternative  and the model-independent approach is to make use of the flavor SU(3) symmetry, which has been argued to work better in charmed baryon decays~\cite{Savage:1989qr,Kohara:1991ug,Verma:1995dk,Chau:1995gk,Sharma:1996sc,Savage:1991wu} and bottomed baryon decays~\cite{Gronau:2013mza,He:2015fwa,He:2015fsa}.

As the experimental precision   is gradually increasing,  the time is ripe to  validate/invalidate  the applicability of the  SU(3) symmetry to charmed baryon decays. The SU(3)  transformation connects  the $\Lambda_c$ with the $\Xi_c$. But at this stage and in the foreseeable future there is no experiment which will  focus on the study on $\Xi_c$ decays. Thus the $\Lambda_c$ decays into various final states especially the ones into a neutron are of great value since they will be the only source for the test of the SU(3) symmetry in charmed-baryon decays.  The motivation of this work is to discuss the roles of  the $\Lambda_c$ decays into a neutron  into the test of SU(3) symmetry and the exploration of  final state interactions, including semileptonic decays, two-body and three-body nonleptonic decays.  All these exclusive decay modes  have not been experimentally measured yet.

This paper is organized as follows. In Sec.~\ref{sec:semileptonic}, the semileptonic $\Lambda_c$ decays are studied.   In Sec.~\ref{sec:two_nonleptonic} and Sec.~\ref{sec:three_nonleptonic}, we will explore the two-body and three-body nonleptonic decays of the $\Lambda_c$, respectively. The last section contains our summary.


\section{Semileptonic $\Lambda_c$ decays}
\label{sec:semileptonic}

We start with the semileptonic $\Lambda_c$ decays.  In the  flavor SU(3) symmetry limit, the charmed baryons are classified according to the SU(3) irreducible representation, namely  as multiplets of the light-quark system: $3\otimes 3 = \bar 3 \oplus 6$. 
The $\Lambda_c$ and $\Xi_{c}$ forms the charmed-baryon anti-triplet in the initial state:
\begin{eqnarray}
 T^a= (\Xi_{c1}^0, -\Xi_{c1}^+, \Lambda_c^+). 
\end{eqnarray}
For the light baryons, we focus on the  SU(3) octet  which is represented by the matrix: 
\begin{eqnarray}
 B^a_b=  \left( \begin{array}{ccc}
    \frac{1}{\sqrt 6} \Lambda^0 + \frac{1}{\sqrt 2} \Sigma^0  & \Sigma^+  & p \\ 
    \Sigma^- &   \frac{1}{\sqrt 6} \Lambda^0 - \frac{1}{\sqrt 2} \Sigma^0  & n \\ 
    \Xi^- & \Xi^0 & -\sqrt{2/3}\Lambda^0  \\ 
  \end{array}\right). 
\end{eqnarray}
The operator responsible  for the transition $c\to q e^+\bar\nu_{e}$ is $[\bar q \gamma^{\mu}(1-\gamma_5) c][\bar \nu_e \gamma_\mu(1-\gamma_5) e]$ with $q=d,s$, which forms an SU(3) anti-triplet in the final state.  Thus the effective Hamiltonian at hadron level  is constructed  as 
\begin{eqnarray}
H_{\rm eff} = a H_a(\bar 3) T^b \bar B^a_b \bar \nu_{e}   e. 
\end{eqnarray}
An implication of the above Hamiltonian is obtained straightforwardly: 
\begin{eqnarray}
{\cal B}(\Lambda_c\to  n  e^+ \nu_e) &=& \frac{3}{2}  \frac{|V_{cd}|^2}{|V_{cs}|^2} {\cal B}(\Lambda_c\to \Lambda e^+ \nu_e).  
\end{eqnarray}
Measurements of the relevant branching fractions  provide  a most straightforward test of the flavor SU(3) symmetry in charmed baryon decays.   
With the most recent data from the BES-III collaboration~\cite{Ablikim:2015prg} 
\begin{eqnarray}\label{eq:Lambdaenu}
{\cal B}(\Lambda_c\to \Lambda e^+ \nu_e)_{\rm BESIII} &=& (3.65\pm 0.38\pm 0.20)\%, 
\end{eqnarray} 
we can obtain  the following result: 
\begin{eqnarray}
{\cal B}(\Lambda_c\to  n  e^+ \nu_e)_{\rm SU(3)} &=& (2.93\pm 0.34)\times 10^{-3}, 
\end{eqnarray} 
which might  be accessible for   BES-III and Belle-II collaborations~\cite{Asner:2008nq,Aushev:2010bq}.

In semileptonic decays, the neutron   can be produced together with a light  pseudo-scalar meson. The  lowest-lying pseudo-scalar meson can be written as
\begin{eqnarray}
 M^a_b=  \left( \begin{array}{ccc}
   \frac{1}{\sqrt 2} \pi^0+  \frac{1}{\sqrt 6} \eta    & \pi^+  & K^+ \\ 
    \pi^- &   - \frac{1}{\sqrt 2} \pi^0+ \frac{1}{\sqrt 6} \eta   & K^0 \\ 
    K^- & \bar K^0 & -\sqrt{2/3}\eta \\ 
  \end{array}\right). 
\end{eqnarray}
In this case, the effective hadronic interaction Hamiltonian is constructed as 
\begin{eqnarray}
 H_{\rm eff} = a [T^a H_a(\bar 3)] (\bar B^c_d M^d_c) \bar\nu_{e} e +  b [T^a \bar B^b_a M^c_b  H_c(\bar 3)]   \bar\nu_{e} e + c [T^a M^b_a \bar B^c_b   H_c(\bar 3)]   \bar\nu_{e} e, 
\end{eqnarray}
where the singlet contribution to $\eta$ has been neglected.  The $a,b,c$ are nonperturbative coefficients.  The above Hamiltonian leads to the expectation: 
\begin{eqnarray}
 {\cal B}(\Lambda_c \to n \overline K^0 e^+\nu_e) =  {\cal B}(\Lambda_c \to p  K^-  e^+\nu_e),
\end{eqnarray}  
which is testable in the near future.  In fact, the above identity holds in the isospin symmetry, whose breaking effect is   much smaller in the charm decays than that of the flavor $SU(3)$ symmetry. In the semi-leptonic decays of $c\to s e^{+}\nu_{e}$, the isospins do not change, $\Delta I=0$. 
It should be stressed here that this identity is applicable to both resonant and non-resonant contributions. 
  
The  branching fraction for the inclusive decay of the $\Lambda_c$ into an electron  has been measured as~\cite{Agashe:2014kda}
\begin{eqnarray}
 {\cal B}(\Lambda_c\to e^++X) = (4.5\pm1.7)\%. 
\end{eqnarray}
Combining the results for the $\Lambda_c\to \Lambda e^+\nu_e$ in (\ref{eq:Lambdaenu}), we may expect:\begin{eqnarray}
 {\cal B}(\Lambda_c \to n \overline K^0 e^+\nu_e) =  {\cal B}(\Lambda_c \to p  K^-  e^+\nu_e)\sim{\cal O}( 10^{-3}).  
\end{eqnarray}  

\section{Two-body nonleptonic $\Lambda_c$ decays } 
\label{sec:two_nonleptonic}
  

For two-body nonleptonic  decays of the $\Lambda_c$, there is no Cabibbo allowed  decay mode  into a neutron.   
Two-body decays into a neutron are either  singly Cabibbo suppressed, 
\begin{eqnarray}
\Lambda_c \to n \pi^+, \;\;\ \Lambda_c \to n \rho^+,  \nonumber
\end{eqnarray}
or doubly Cabibbo suppressed, 
\begin{eqnarray}
 \Lambda_c\to n K^+, \;\;
 \Lambda_c\to n K^{*+}. 
\end{eqnarray}  

The nonleptonic $\Lambda_c$ decays are induced by the operators $[\bar s c][\bar ud]$ for the Cabibbo-allowed mode   and $[\bar d c][\bar ud]$ for the Cabibbo-suppressed mode. These operators can be decomposed into irreducible representations of flavor SU(3). For instance, 
\begin{eqnarray}
( \bar s c )(\bar ud) = {\cal O}_6 + {\cal O}_{\overline {15}}, 
\end{eqnarray}
with 
\begin{eqnarray}
{\cal O}_6 = \frac{1}{2} [(\bar sc )(\bar ud)-(\bar u c )(\bar s d)], \nonumber\\
{\cal O}_{\overline {15}} = \frac{1}{2} [(\bar sc )(\bar ud)+(\bar u c )(\bar s d)]. 
\end{eqnarray}
Perturbative QCD corrections give rise to an enhancement of the coefficient for the ${\cal O}_6$ over the coefficient for the ${\cal O}_{\overline {15}}$ by~\cite{Gaillard:1974nj,Altarelli:1974exa}
\begin{eqnarray}
 \left[\frac{\alpha_s(m_b)}{\alpha_s(m_W)}\right]^{18/23}  \left[\frac{\alpha_s(m_c)}{\alpha_s(m_b)}\right]^{18/25}  \sim 2.5. 
\end{eqnarray}
If this is valid, then one has 
\begin{eqnarray}
 H_{\rm eff} = e H^{ab}(6) T_{ac} \bar B^c_d M^d_b + f H^{ab}(6) T_{ac} M^c_d \bar B^d_b + g H^{ab}(6) \bar B^c_a M^d_b T_{cd}, \label{eq:hamil_two_non}
\end{eqnarray}
with $H^{22}(6)=1$ for Cabibbo-allowed modes,  $H^{23}(6)= H^{32}(6)= -2\sin(\theta_C)$ for singly Cabibbo-suppressed modes, and $H^{33}(6)= +2 \sin(\theta_C)^2$ for doubly-Cabibbo-suppressed modes, where $\theta_{C}$ is the Cabibbo angle,  and 
\begin{eqnarray}
 T_{ab}=\epsilon_{abc}T^c. 
\end{eqnarray}
The coefficients $e,f,g$ are the nonperturbative amplitudes.

Using Eq.~(\ref{eq:hamil_two_non}), we find that for the doubly-Caibbo-suppressed modes:
\begin{eqnarray}
 {\cal B}(\Lambda_c\to nK^+) =  {\cal B}(\Lambda_c\to pK^0). 
\end{eqnarray}
For the singly-Cabibbo-suppressed modes,  we have the decay amplitudes, 
\begin{eqnarray}
 {\cal A}(\Lambda_c\to n\pi^+) =  \sqrt{2}{\cal A}(\Lambda_c\to p\pi^0)= (2f+2g)\sin(\theta_C), 
\end{eqnarray}
which implies the relation: 
\begin{eqnarray}
 {\cal B}(\Lambda_c\to n\pi^+) =  2{\cal B}(\Lambda_c\to p\pi^0). 
\end{eqnarray}

Furthermore, we have  the amplitudes for Cabibbo-allowed modes: 
\begin{eqnarray}
 {\cal A}(\Lambda_c\to \Lambda\pi^+) &=& \frac{1}{\sqrt 6} ( -2e-2f -2g),\\
 {\cal A}(\Lambda_c\to \Sigma^0\pi^+) &=& \frac{1}{\sqrt 2} ( -2e+2f +2g),\\
 {\cal A}(\Lambda_c\to p\overline K^0) &=& -2e. 
\end{eqnarray}
Thus we can derive  the sum rule that can be experimentally examined: 
\begin{eqnarray}
 {\cal B}(\Lambda_c\to n\pi^+)=  \sin^2(\theta_C)\left[ 3{\cal B}( \Lambda_c\to \Lambda\pi^+) +  {\cal B}(\Lambda_c\to \Sigma^0\pi^+)- {\cal B}(\Lambda_c\to p\overline K^0) \right] . 
\end{eqnarray}

The recent  BES-III data~\cite{Ablikim:2015flg}  implies 
\begin{eqnarray}
 {\cal B}(\Lambda_c\to n\pi^+) =  \sin^2(\theta_C)\left[3\times 1.24\% + 1.27\% - 3.04\% \right]  \sim 0.9\times 10^{-3}. 
\end{eqnarray}
Measurements in future by BES-III will be able to validate/invalidate the dominance of the sextet assumption in the effective operator.

\section{Three-body nonleptonic $\Lambda_c$ decays}
\label{sec:three_nonleptonic}

Compared to two-body decays, three-body $\Lambda_c$ decays are more involved, since first they can proceed via quasi-two-body process and the non-resonant decays and secondly there are a number of independent amplitudes in SU(3) symmetry.  In the following  we consider the $NK\pi$ system in the isospin limit: 
\begin{eqnarray}\label{eq:isos}
 |p\overline K^0\pi^0\rangle &=& |\frac{1}{2}\frac{1}{2}\rangle  |\frac{1}{2}\frac{1}{2}\rangle |1 0\rangle  = |1 1\rangle |10\rangle   =  \frac{1}{\sqrt2} |21\rangle + \frac{1}{\sqrt2} |11\rangle^{(1)}, \\ 
  |p K^-\pi^+\rangle &=&  |\frac{1}{2}\frac{1}{2}\rangle  |\frac{1}{2}-\frac{1}{2}\rangle |1 1\rangle  =  \left(\frac{1}{\sqrt2} |10\rangle + \frac{1}{\sqrt2} |00\rangle\right) |11\rangle   
   =  \frac{1}{2} |21\rangle -\frac{1}{2} |11\rangle^{(1)} + \frac{1}{\sqrt2} |11\rangle^{(2)},\\ 
  |n\overline K^0\pi^+\rangle &=&  |\frac{1}{2}-\frac{1}{2}\rangle  |\frac{1}{2}\frac{1}{2}\rangle |1 1\rangle  =  \left(\frac{1}{\sqrt2} |10\rangle - \frac{1}{\sqrt2} |00\rangle\right) |11\rangle  =  \frac{1}{2} |21\rangle -\frac{1}{2} |11\rangle^{(1)} - \frac{1}{\sqrt2} |11\rangle^{(2)},\label{eq:isos1}
\end{eqnarray}
where the superscripts $(1)$ and $(2)$ are isospin states from $(1-1)$ and $(0-1)$ couplings, respectively, which are independent with each other.
Since the Hamiltonian of the $c\to s\bar du$ transition has $\Delta I=1$, and the isospin of $\Lambda_{c}$ is zero, we can derive the decay amplitudes from the above decompositions: 
\begin{eqnarray}\label{eq:isospin12}
{\cal A}(\Lambda_c\to p\overline K^0\pi^0) &=& \frac{1}{\sqrt 2} {\cal A}^{(1)}, \nonumber\\
{\cal A}(\Lambda_c\to p  K^-\pi^+) &=& -\frac{1}{ 2} {\cal A}^{(1)}+  \frac{1}{\sqrt 2} {\cal A}^{(2)}, \nonumber\\
{\cal A}(\Lambda_c\to n  \overline K^0\pi^+) &=& -\frac{1}{ 2} {\cal A}^{(1)}-  \frac{1}{\sqrt 2} {\cal A}^{(2)}.
\end{eqnarray}
The above amplitudes lead to the sum rule
\begin{eqnarray}\label{eq:sumrule}
 \sqrt{2}{\cal A}(\Lambda_c\to p\overline K^0\pi^0) + {\cal A}(\Lambda_c\to p  K^-\pi^+) + {\cal A}(\Lambda_c\to n  \overline K^0\pi^+) =0. 
\end{eqnarray}
Note that the isospin amplitudes in eq.(\ref{eq:isospin12}) can be changed if we firstly couple the $K\pi$ states from eq.(\ref{eq:isos}-\ref{eq:isos1}), but the sum rule in eq.(\ref{eq:sumrule}) still holds.



Measurements of branching ratios of the three channels are able to determine the two  amplitudes, and in particular investigate the relative strong phases between the two independent decay amplitudes. These phases arise from the final state interactions since if factorization works, the two independent  amplitudes are real with vanishing phases at leading order.  These amplitudes including phases  can provide the essential inputs for the analysis of  nonleptonic decays  into other baryons like $\Lambda$.


%

\begin{figure}[phtb]
\begin{center}
\includegraphics[scale=1]{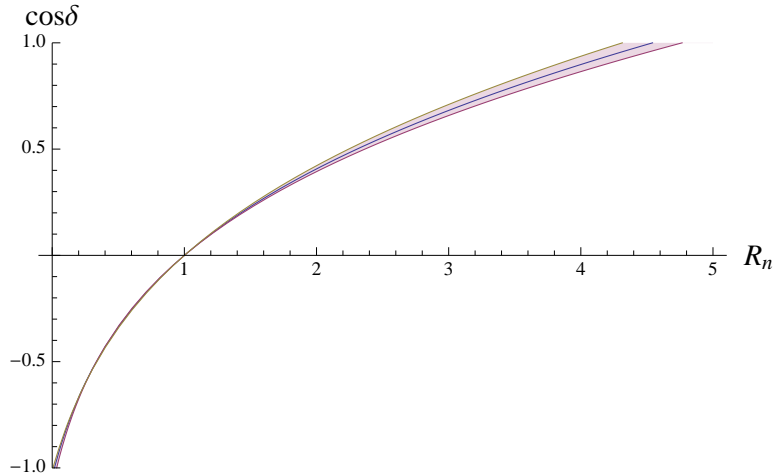}
\end{center}
\caption{Correlation  between $\cos\delta$ and $R_{n}$, with $\delta$ as the strong phase difference in Eq.~(\ref{eq:phase}) and $R_{n}$ as the ratio of branching fractions in Eq.~(\ref{eq:ratio}). }\label{fig:cosdelta}
\end{figure}

From eq.(\ref{eq:isospin12}), we define the relative strong phase, $\delta$, between $\mathcal{A}^{(1)}$ and $\mathcal{A}^{(2)}$ :
\begin{equation}
{\mathcal{A}^{(2)}\over\mathcal{A}^{(1)}}=\left|{\mathcal{A}^{(2)}\over\mathcal{A}^{(1)}}\right|e^{i\delta}.\label{eq:phase}
\end{equation}
Then the branching fractions can be expressed as
\begin{eqnarray}
\mathcal{B}(\Lambda_{c}\to p\overline K^{0}\pi^{0})&=&{1\over2}\left|\mathcal{A}^{(1)}\right|^{2},
\nonumber\\
\mathcal{B}(\Lambda_{c}\to p K^{-}\pi^{+})&=&{1\over4}\left|\mathcal{A}^{(1)}\right|^{2}+{1\over2}\left|\mathcal{A}^{(2)}\right|^{2}-{1\over\sqrt2}\left|\mathcal{A}^{(1)}\right|\left|\mathcal{A}^{(2)}\right|\cos\delta,
\\
\mathcal{B}(\Lambda_{c}\to n\overline K^{0}\pi^{+})&=&{1\over4}\left|\mathcal{A}^{(1)}\right|^{2}+{1\over2}\left|\mathcal{A}^{(2)}\right|^{2}+{1\over\sqrt2}\left|\mathcal{A}^{(1)}\right|\left|\mathcal{A}^{(2)}\right|\cos\delta,
\nonumber
\end{eqnarray}
where we consider  the relative strong phase to understand the final state interaction, and neglect the phase spaces which are actually integrated in the three-body decays.
Hence
\begin{equation}
\cos\delta=\frac{\mathcal{B}(n\overline K^{0}\pi^{+})-\mathcal{B}(p K^{-}\pi^{+})}{2\sqrt{\mathcal{B}(p\overline K^{0}\pi^{0})\left(\mathcal{B}(p K^{-}\pi^{+})+\mathcal{B}(n\overline K^{0}\pi^{+})-\mathcal{B}(p\overline K^{0}\pi^{0})\right)}}.
\end{equation}
Defining
\begin{equation}
R_{p}={\mathcal{B}(\Lambda_{c}\to p\overline K^{0}\pi^{0})
\over\mathcal{B}(\Lambda_{c}\to p K^{-}\pi^{+})},~~~~~R_{n}={\mathcal{B}(\Lambda_{c}\to n\overline K^{0}\pi^{+})
\over\mathcal{B}(\Lambda_{c}\to p K^{-}\pi^{+})}, \label{eq:ratio}
\end{equation}
we have
\begin{equation}
\cos\delta={R_{n}-1 \over 2\sqrt{R_{p}(1+R_{n}-R_{p})}}.
\end{equation}
From the recent measurement by BESIII \cite{Ablikim:2015flg}, $R_{p}=0.64\pm0.06$. Then $\cos\delta$ can be obtained once the $R_{n}$ is measured. The relation between $\cos\delta$ and $R_{n}$ is shown in Fig.\ref{fig:cosdelta}.
Since $-1\leq\cos\delta\leq1$, we have $0.017\leq R_{n}\leq 4.54$, and then the branching fraction of $\Lambda_{c}\to n\overline K^{0}\pi^{+}$ is obtained as,
\begin{eqnarray}
0.04\%\leq \mathcal{B}(\Lambda_{c}\to n\overline K^{0}\pi^{+})_{\rm Belle}\leq33\%, \\
0.035\%\leq \mathcal{B}(\Lambda_{c}\to n\overline K^{0}\pi^{+})_{\rm BESIII}\leq28\%.
\end{eqnarray} 
As we can see that this constraint is rather loose, thus the   experimental measurements are  requested.

\section{Summary}
\label{sec:summary}

Unlike the bottom hadron decays where the momentum transfer is typically large enough to ensure the perturbation theory in QCD,   charmed meson and baryon decays are very difficult to understand. Due to the limited energy release, the factorization scheme    based on the expansion of $1/m_c$ and $1/E$   is not always valid. 
Flavor SU(3) symmetry is a powerful  tool to analyze the charmed baryon decays, which has been argued to work better than in charmed meson decays, however its validity  has to be experimentally examined.  Since there is not much  data on $\Xi_c$ decays, exclusive $\Lambda_c$ decays into a neutron are  essential  for the test of flavor symmetry and investigating  final state interactions in charmed baryon decays. 

In this work, we have  discussed  the  roles of  the  exclusive $\Lambda_c$ decays into a neutron in testing the flavor symmetry  and  final state interactions. We found that the    semileptonic decays into a neutron provide the most-straightforward way to explore the flavor SU(3) symmetry.  Two-body nonleptonic decays are capable to examine the assumption of the sextet dominance mechanism. While   three-body non-leptonic decays into a neutron are of great interest to explore the final state interactions in $\Lambda_c$ decays.  All these decay modes have not been experimentally observed to date.



 \section*{\it Acknowledgements:}
The authors are very  grateful to Xiao-Gang He, Lei Li, Xiao-Rui Lyu, and Yang-Heng Zheng  for  enlightening discussions. This work is supported in part  by National  Natural  Science Foundation of China under Grant  No.11575110, 11521505, 11347027, 11505083, 11235005 and 11375208,  Natural Science Foundation of Shanghai under Grant  No.15DZ2272100 and No.15ZR1423100,  by the Open Project Program of State Key Laboratory of Theoretical Physics, Institute of Theoretical Physics, Chinese  Academy of Sciences, China (No.Y5KF111CJ1), and  by   Scientific Research Foundation for  Returned Overseas Chinese Scholars, State Education Ministry.


\end{document}